# The riddle of the plant vacuolar sorting receptors

***Running head***: Masclaux et al., Plant vacuolar sorting receptors

Frédéric G. Masclaux[2], Jean-Philippe Galaud[1] and Rafael Pont-Lezica[1*]

[1)] Surfaces Cellulaires et Signalisation chez les Végétaux, UMR 5546, CNRS-Université Paul Sabatier, BP 17 Auzeville, 31326 Castanet-Tolosan, France.
[2)] Present address: Université de Genève, Laboratoire de Génétique végétale, Sciences III 30, Quai Ernest-Ansermet, CH-1211 Genève 4, Switzerland

* Corresponding author
Telephone: +33 5 6219 3516
Fax: +33 5 6219 3502
Mail: **lezica@scsv.ups-tlse.fr**

## Summary

Proteins synthesized on membrane-bound ribosomes are sorted at the Golgi apparatus level, for delivery to various cellular destinations: the plasma membrane or the extra-cellular space, and the lytic vacuole or lysosome. Sorting involves the assembly of vesicles, which preferentially package soluble proteins with a common destination. The selection of proteins for a particular vesicle type involves the recognition of proteins by specific receptors, such as the vacuolar sorting receptors for vacuolar targeting. Most eukaryotic organisms have one or two receptors to target proteins to the lytic vacuole. Surprisingly, plants have several members of the same family, seven in *Arabidopsis thaliana*. Why plants have so many proteins to sort soluble proteins to their respective destinations? The presence of at least two types of vacuoles, the lytic and the storage one, seems to be a partial answer. In this review we analyze the last experimental evidence supporting the presence of different subfamilies of plant vacuolar sorting receptors.

**Abbreviations:** BP-80, binding protein of 80 kDa; CCV, clathrin-coated vesicles; ELP, epidermal growth factor-like protein; ER, endoplasmic reticulum; GGA: Golgi colocalized γ-ear-containing ARF-binding proteins; MPR, mannose 6-phosphate receptor; MRL1p, mannose 6-phosphate receptor like protein; PSV, protein storage vacuole; TGN, trans-Golgi network; VSR, vacuolar sorting receptor.

## Introduction

In eukaryotic cells, proteins synthesized on membrane-bound ribosomes are transported through the Golgi apparatus and are sorted for delivery to various cellular destinations. The Golgi apparatus has a central role in the modification and sorting of proteins. It is a very dynamic compartment made of distinct cisternae with a cis-trans organization that are involved in vacuolar protein sorting (Hillmer et al., 2001; Ward and Brandizzi, 2004). The classical view has been that the traffic bifurcates at the TGN: to the constitutive default pathway, delivering the proteins to the cell surface; or to a selective pathway, sorting proteins to the endosomal membrane system (Mellman and Warren, 2000).

The endosomal membrane system is characterized by the presence of acid hydrolases and is known as lysosome/endosome in mammalian cells, or vacuole in yeasts and plants. The sorting at the TGN is mediated by transmembrane receptors recruiting cargo proteins in the lumen side. Until 2001, it was assumed that the hetrotetrameric clathrin coat adaptator AP-1 was responsible for clathrin-coated vesicle (CCV) formation but a new family of monomeric clathrin coat proteins, the Golgi colocalized γ-ear-containing ARF-binding proteins (GGA) was identified. It was proposed that these GGA proteins, in association with clathrin, are involved in the creation of vesicles destined to late endosome (Black and Pelham, 2000; Puertollano et al., 2001a; Puertollano et al., 2001b). CCV deliver the receptor-ligand complexes to the lysosomal/vacuolar system (Ghosh et al., 2003). Conserved sequences at the receptor's cytosolic domain allow the recognition of the protein complexes delivering the vesicles to the correct destination.

The transmembrane receptors involved in the sorting process are known as vacuolar sorting receptors (VSR). The soluble hydrolases of the mammalian lysosome are marked for delivery to this organelle by the addition of mannose 6-phosphate to their glycans. Two mannose 6-phosphate receptors (MPR) were identified by their ability to bind mannose 6-phosphate, the 46 kDa cation-dependent MPR, and the 300 kDa cation-independent MPR (Kornfeld, 1992). In addition to the delivery of acid hydrolases to the endosomes, the cation-independent MPR has been implicated in various physiological processes, including the internalization of insulin-like growth factor (Gosh et al., 2003). In contrast, in yeast and plants, the vacuolar targeting signals are part of the peptidic sequence itself (Robinson and Hinz, 1997). The only known receptor from yeast VPS10p recognizes a specific amino acid sequence of carboxypeptidase Y (Marcusson et al., 1994; Conibear and Stevens, 1998; Gerrard et al., 2000a, 2000b). A membrane protein from yeast, mannose 6-phosphate receptor like protein (MRL1p), distantly related to mammal MPR has been recently identified and may be a new VSR (Whyte and Munro, 2001), but clear evidence of this function is missing. It has been recently found that sortilin, a VPS10p-like receptor, is also present in animal cells and seems to be implicated in Golgi-endosome transport (Nielsen et al., 2001). Surprisingly, sortilin seems to be also involved in neuronal death (Nykjaer et al., 2004). In plants, targeting to the lytic vacuole is mediated by an integral membrane protein named BP-80 (binding protein of 80 kDa). This receptor identified in pea and *Arabidopsis* recognizes the NPIR motif at the N-terminal of aleurain and sporamin, and has no sequence homology to any other known sorting receptor in eukaryotes (Kirsch et al., 1994; Paris et al., 1997; Humair et al., 2001). They are type I integral membrane proteins with three repeats of an epidermal growth factor motif close to the transmembrane domain that may stabilize the ligand-receptor complexes (Watanabe et al., 2004). In addition, the N-terminal domain contains a conserved protease associated region. The cytosolic C-terminus carries an YXXL sequence, a potential

Tyr-based motif, which interacts with the *Arabidopsis* μA-adaptin which is a subunit of a putative AP-complex located at the trans-Golgi Network (Happel et al., 2004).

## The *Arabidopsis* VSR multigenic family

The availability of *Arabidopsis* complete genome revealed the presence of seven members of the family in this plant. Some members of the *Arabidopsis* genes were identified, cloned, and given different names (table I). BP-80 came from the size of the protein binding the vacuolar targeting sequence from aleurain, and all the genes were named accordingly (Paris et al. 1997; Hadlington and Denecke, 2000**)**. Those proteins were also named as ELP (Epidermal growth factor-Like Protein) in agreement to the presence of specific protein motifs (Ahmed et al., 1997; Laval et al., 1999). The last classification, based on the function of the proteins, seems the more convenient and will be used in this review, even if that function has not been experimentally proved for all the members of the family (Paris et al., 1997; Shimada et al., 2003).

Table I. Equivalence between the different names given to the *Arabidopsis* members of vacuolar sorting receptors. AtVSR: *Arabidopsis thaliana* Vacuolar Sorting Receptor (Shimada et al., 2003); AGI: locus of the gene in the genome given by the *Arabidopsis* Genome Initiative (2000); BP80: binding protein of 80 kDa (Hadlington and Denecke, 2000); ELP: Epidermal growth factor-Like Protein (Ahmed et al., 1997; Laval et al., 1999).

| VSR | AGI | BP80 | ELP |
|---|---|---|---|
| AtVSR1 | At3g52850 | Atbp80b | AtELP1 |
| AtVSR2 | At2g30290 | Atbp80c | AtELP4 |
| AtVSR3 | At2g14740 | Atbp80a’ | AtELP2a |
| AtVSR4 | At2g14720 | Atbp80a | AtELP2b |
| AtVSR5 | At2g34940 | Atbp80d | AtELP5 |
| AtVSR6 | At1g30900 | Atbp80e | AtELP6 |
| AtVSR7 | At4g20110 | Atbp80f | AtELP3 |

The main question we address here is why plants have so many VSR when other eukaryotes have only one or two? The common answer is that plants have a more complex vacuolar system than other eukaryotes. At present, two clearly different types of vacuoles have been identified in plants: the lytic vacuoles equivalent to the endosome/lysosome from mammals and the yeast vacuole, and the storage vacuole present mainly in seed reserve tissue (Paris et al., 1996; Robinson and Hinz, 1997; Jauh et al., 1998). However, most of the excellent reviews on the area considered the family of AtVSRs as mainly involved in the transport to the lytic vacuole (Beevers and Raikhel, 1998; Neuhaus and Rogers, 1998; Paris and Neuhaus, 2002). The role of the pea receptor (BP80) as a vacuolar sorting receptor was reported by the use of a yeast mutant knocked out on VPS10p. The pea receptor was able to target a reporter green fluorescent protein construction carrying the aleurain sorting N-terminal signal to the lytic vacuole, but did not restore the vacuole localization of yeast carboxypeptidase Y (Humair et al., 2001). Since yeast contains only one type of vacuole, the lytic one, which is considered as the default pathway for heterogeneous expressed membrane proteins, the results of such experiments only partially support the in vivo function (Takegawa et al., 2003).

A controversial report showed that a pumpkin protein closely related to the *Arabidopsis* VSRs, PV72 was an abundant protein in precursor-accumulating vesicles from maturing pumpkin seeds (Shimada et al., 1997). PV72 was able to bind peptides derived from a 2S albumin as well as the precursor protein, a protein found in the storage vacuoles. Since the *Arabidopsis* VSRs had been isolated from CCVs delivering proteins to the lytic vacuole, the involvement of such receptors in the sorting of storage proteins was discarded. The diversity of species and tissues used in those studies made difficult the interpretation of the results obtained so far. Recently, a tagged mutant for the *AtVSR1* gene, a close homologue to pumpkin PV72, was shown to secrete reserve proteins into the extracellular space. However, some storage proteins were still correctly targeted to the PSV. This may indicate specificity between cargo proteins and the corresponding VSR, or that other storage proteins were targeted via the alternative ER-derived compartments. In wild type seeds, the AtVSR1 protein binds the C-terminal peptide of the 12S globulin in a $Ca^{2+}$ dependent manner (Shimada et al, 2003). Curiously, a fusion protein containing the pumpkin PV72 lumenal domain with an ER-retention signal introduced in *Arabidopsis*, retained NPIR-containing proteases in the ER, but not other vacuolar proteins (Watanabe et al., 2004). This indicates that PV72 may recognize proteins to be sorted to the lytic or the storage vacuole. It is interesting to point out that the lytic enzymes were normally sorted to the lytic vacuoles in the *atvsr1* mutant (Shimada et al, 2003). It clearly showed that AtVSR1 is involved in the transport of some soluble proteins to the storage vacuoles, but not in the sorting of proteins to the lytic compartment. This brings another question. How are degraded the miss-sorted reserve proteins in the *atvsr1* mutant? We do not know the answer; however we can advance some possibilities. First the reserve proteins may be internalized to the lytic vacuole by an endocytic pathway not well known in plants. Second, they may be degraded in the extracellular space by secreted proteases. Recent evidences on the cell wall proteome in *Arabidopsis* indicate that an important number of proteases are normally present in the cell wall of rosettes and other organs (Boudart et al, 2005). In a recent work, Jolliffe et al. (2004) showed that the transport of proricin and pro 2S albumin to the protein storage vacuole in castor bean involves members of the VSR family. These data support the work of Shimada et al. (2003) and indicate that our understanding of the role of VSR in protein sorting is still incomplete. To add a new level of complexity, we have to take into consideration that in seeds, the protein storage vacuole includes a membrane-bound compartment of different composition and function (Jiang et al., 2001, Nishizawa et al., 2003). It was proposed that this globoid cavity represents a lytic like compartment inside the PSV. These results are very exciting since we can imagine that, at least in seeds, some members of vacuolar receptors can deliver proteins in a specific compartment of the protein storage vacuole.

It should be noted that in plants, in addition to the classical transport via Golgi pathway, some proteins and other molecules can apparently be stored for shorter or longer periods of time in endoplasmic reticulum-derived compartments. In some cases these ER derived compartments travel to and are incorporated into vacuoles. Plant cells appear to have flexibility in using the ER to assemble storage organelles. So far, ER-derived storage organelles such as protein bodies and oil bodies have been described primarily in seeds (Chrispeels and Herman, 2000). At present, it is not known if membrane receptors are involved in the recruitment of proteins for ER-derived vesicles.

## The inhibition of VSR gene expression affects germination.

In yeast, the knockout of the *VPS10* gene does not show a visible phenotype; only some proteases are miss-sorted but the cell growth perfectly well. A similar situation seems to occur in plants, where tagged mutants had not been particularly useful for the understanding of protein trafficking (Zouhar et al., 2004). The only visible phenotype was obtained by an antisense construction blocking the expression of all the members of the *AtVSR* family, where most of the lines failed to germinate (Laval et al., 2003). Germination is completely dependent on stored reserves since the embryo has not yet acquired autotrophy. In this particular situation are all the *Arabidopsis* VSRs involved or only some of them? Recently it was shown that germination depends on stored and not on neosynthesized mRNA (Rajjou et al., 2004). Subsequent growth of the seedling needs the newly synthesized mRNA. An expression study of the *AtVSR* family indicates that only three of the genes are transcribed in dry seeds, namely *AtVSR1*, *3* and *4* and after imbibition. The others genes are expressed only in the growing seedling and later (Laval et al., 2003; Nakabayashi et al., 2005). It seems clear that germination depends not only on the reserve proteins from the storage vacuole, but also on the proteases from the lytic vacuole.

## Phylogenetic trees suggest different functions

The N-terminal moiety of VSR proteins binds the cargo proteins that will be targeted to a particular destination. The alignment of the central part of the Arabidopsis VSR gives a Phylogenetic tree showing three main groups (Paris and Neuhaus, 2002; Shimada et al., 2003). On the other hand, the C-terminus region contains the receptor sequences that should bind to the cytosolic proteins involved in vesicular traffic from the TGN to the various pre-vacuolar compartments from the plant cell. We have constructed a phylogenetic tree aligning the C-terminal region of the AtVSR family plus some other close related proteins found in databases (Fig. 1). This tree gives the same three groups obtained previously by alignment of the N-terminal moiety. Two *Arabidopsis* members (AtVSR3 and 4) and two rice sequences are in the same branch as the pea PB80, which was found to be involved in the sorting to the lytic vacuole (Paris et al., 1997; Humair et al., 2001). A second group assembles the pumpkin PV72, AtVSR1 and 2, plus a wheat and *Vigna* members. This branch comprises VSR members involved in targeting storage proteins to the storage vacuole (Shimada et al, 1997; Shimada et al., 2003). The open question is the putative function of the third branch. Those three members are normally expressed in almost all the tissues (Laval et al., 2003); even if the proteins have not been isolated yet. Based on the expression data of all the *AtVSR* genes in various organs and at different developmental stages, a cluster analysis was performed (not shown); this tree is quite similar to the one shown in Fig. 1.

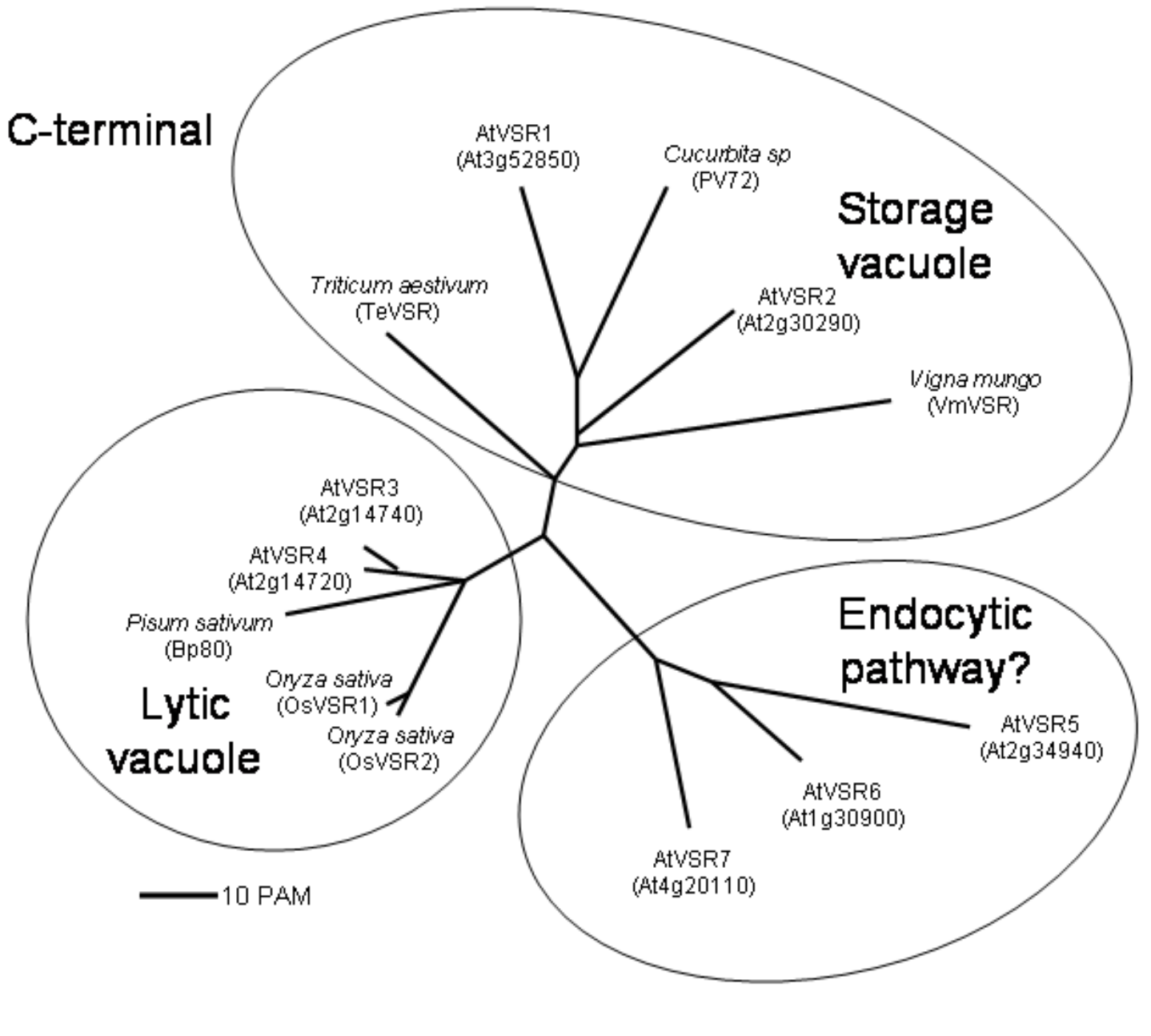

**Figure 1.** Phylogenetic tree of VSR proteins of *Arabidopsis*, pea, pumpkin, rice, wheat and black gram. The C-terminal region between Trp-538 and Asp-624 of AtVSR4 was aligned with the corresponding regions of the other proteins from ProDom database and tools (Servant et al., 2002).

It has been shown by immuno-electron microscopy that the pea BP-80 and AtVSR1 were found in CCV, at the *trans* face of the Golgi, and in prevacuoles (Kirsh et al., 1994; Paris et al, 1997, Amhed et al., 1998). This shows clearly that the VSR is not found in the final destination, but only in the compartments where the sorting is made and in intermediate vesicles. Using antibodies raised against peptides shared by all members of the family, it was shown that a VSR-like protein was present in a fraction enriched in plasma membrane proteins from *Arabidopsis* (Laval et al., 1999). Little is known on the endocytic pathway in plants. This pathway had been extensively studied in mammals, where the calcium independent-MPR can sort surface proteins to the endosome as well as to target soluble proteins to the lysosome. We propose here, that the third branch may group plant proteins involved in the endocytic pathway. Another putative role for the members of that branch may be their involvement in the sorting of surface-destined cargo proteins in polarized cells. Animal polarized cells will deliver surface proteins to the apical or basolateral surface into distinct carriers at the TNG (Traub and Kornfeld, 1997; Bonifacio and Traub, 2003; Gosh et al., 2003). It is known that membrane proteins, like the auxin transporter PIN1, are secreted to the basolateral surface of most cells (Palme and Galweiler, 1999). Auxin transport inhibitors and the vesicle trafficking inhibitor, brefeldine A, blocked PIN 1 cycling (Geldner et al., 2001**;** Petrasek et al., 2003). It seems possible that the secretion of proteins into specific extracellular areas does not follow the default pathway, and may need sorting membrane proteins. Figure 2, gives a schematic view of the state of the art, with some experimentally proved roles, plus some new hypothesis that may stimulate research in the area. It seems clear that the targeting of proteins to different locations within the plant cell is more complex than supposed previously and need intensive research.

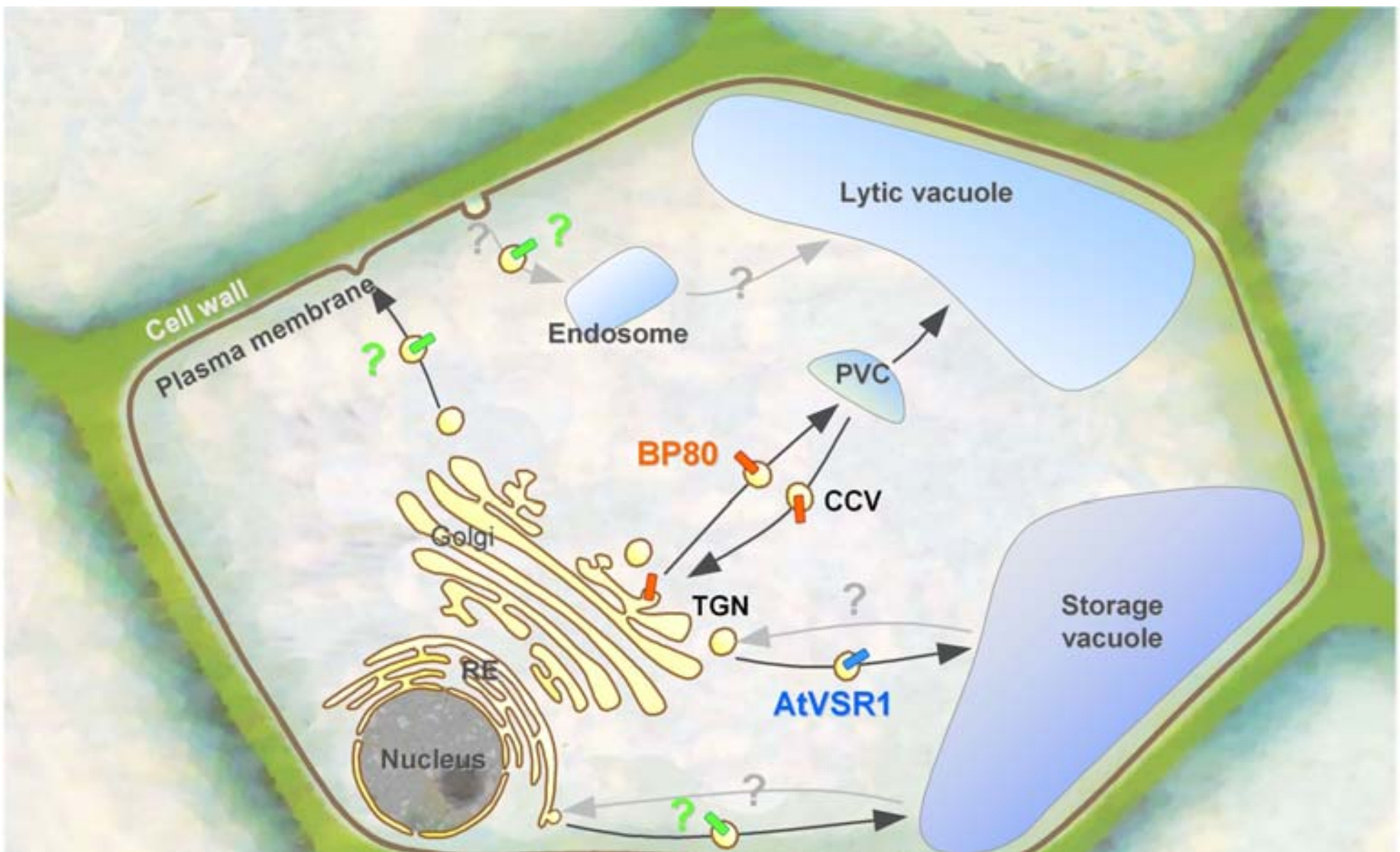

**Figure 2.** Schematic representation of the different pathways to the lytic, storage, and endocytic routes. BP80: the first VSR isolated from pea and involved in the targeting of proteins to the lytic vacuole (Paris et al, 1997; Humair, 2001). AtVSR1: *Arabidopsis thaliana* Vacuolar Sorting Receptor 1, implicated in the targeting of proteins to the storage vacuoles (Shimada et al., 2003). PVC: pre-vacuolar compartment. CCV: clathrin-coated vesicle. TGN: trans-Golgi network. ER: endoplasmic reticulum.